# Influences of $N_2$ flow rate on the superconducting properties and dimensionality of epitaxial TiN thin films*

CHEN Shilong[1,2], LIU Yuchuan[1,2], SUN Yihui[1,2], WANG Xiaoni[2], PENG Wei[2], WU Yu[2,†], MU Gang[2,†], LIN Zhirong[1,2,†]

1. School of Microelectronics, Shanghai University, Shanghai 201800, China
2. Shanghai Institute of Microsystem and Information Technology, Chinese Academy of Sciences, Shanghai 200050, China

**Abstract** TiN thin films have received great attention in the field of superconducting quantum computing due to their low loss characteristics derived from stable chemical and thermal properties, as well as excellent electrical conductivity. In this study, TiN thin films with a thickness of 35 nm were epitaxially grown on MgO substrates using the DC reactive magnetron sputtering technique. The crystal structure, film thickness, and surface morphology of the thin films were characterized. The films exhibit a very flat surface with a root mean square roughness of only a few tenths of a nanometer. The influences of nitrogen ($N_2$) flow rate on the surface morphology, superconducting transition temperature ($T_c$), residual resistivity ratio (RRR), normal-state resistivity ($\rho_n$), upper critical field, and superconducting dimensionality were investigated systematically. The experimental results show that the TiN film deposited with a $N_2$ flow rate of 1.5 sccm exhibits optimal performance, including the highest $T_c$ (4.93 K), the largest RRR, the lowest $\rho_n$, and the lowest slope of the out-of-film upper critical field near $T_c$. Moreover, it is found that the background vacuum inside the growth chamber has a significant impact on the superconducting properties of TiN films, and the degradation of vacuum degree can lead to the decrease of $T_c$. The upper critical field reveals a significant anisotropic feature. Importantly, the in-film upper critical field data demonstrates that the optimization of sample quality leads to deviation from two-dimensional superconducting behavior, which provides a new perspective on the influencing factors of superconducting dimensionality beyond the thickness of films. The present results provide useful references for a comprehensive understanding of the

* The paper is an English translated version of the original Chinese paper published in Acta Physica Sinica. Please cite the paper as: **CHEN Shilong, LIU Yuchuan, SUN Yihui, WANG Xiaoni, PENG Wei, WU Yu, MU Gang, LIN Zhirong, Influences of $N_2$ flow rate on the superconducting properties and dimensionality of epitaxial TiN thin films.** *Acta Phys. Sin.*, 2026, 75(11): 110703. DOI: 10.7498/aps.75.20260155

†Corresponding Authors. E-mail: wuyu@mail.sim.ac.cn, mugang@mail.sim.ac.cn and zrlin@mail.sim.ac.cn

relationship between growth parameters and the superconducting properties of thin films.

# 1 Introduction

Superconducting thin films serve not only as the material foundation for device applications but also as an ideal platform for investigating low-dimensional superconducting phenomena[1-6]. Nitride superconducting materials are widely employed in superconducting electronic devices due to their favorable chemical stability, processability, and superconducting properties[7-9]. Among these, TiN films exhibit excellent oxidation resistance, thermal stability, and good electrical conductivity, demonstrating significant potential for electronic device applications. As superconductors, TiN films feature low microwave loss[10] and high kinetic inductance[11]. These characteristics enable their broad application in various quantum technology fields, including superconducting qubit circuits[12,13], single-photon detectors[14], superconducting amplifiers[15,16], and quantum phase slip devices[17]. In recent years, the application of TiN films in superconducting microwave resonators[18-21] has attracted increasing attention. Results reported by Narayan et al.[22] and Vispute et al.[23] indicate that the quality factor of microwave resonators fabricated from TiN films can exceed $2\times10^6$.

TiN films are typically prepared using various methods, such as magnetron sputtering[24], pulsed laser deposition[25,26], and molecular beam epitaxy[27]. Since the superconducting properties of thin films directly influence their loss characteristics in quantum circuits, investigating how growth conditions affect the superconducting performance of TiN films holds significant practical value. Previous studies by our group have demonstrated that the substrate type significantly influences the crystal orientation of TiN films[28]. Notably, MgO substrates share the same NaCl-type crystal structure as TiN and exhibit a small lattice mismatch. This facilitates the growth of high-quality (100)-oriented TiN films[24]. Although reports exist on the impact of growth parameters on film properties[29], systematic research linking TiN film growth conditions to their superconducting behavior, particularly regarding critical magnetic field, anisotropy, and coherence length, remains lacking.

Furthermore, as dimensions decrease, superconducting materials exhibit low-dimensional characteristics distinct from bulk materials. Examples include the superconductor-insulator transition[30-33], quantum phase slips[34-38], quantum Griffiths singularity[39-43], Ising superconductivity[44-46], and quantum metal states[47-50]. In this field, the transition from three-dimensional (3D) to two-dimensional (2D) superconductivity is typically attributed to the evolving relationship between film thickness and superconducting coherence length, with less attention paid to other influencing factors. Conducting systematic studies on film growth conditions and sample quality helps further reveal the evolution laws of superconducting dimensionality.

This work systematically explores the regulatory effect of $N_2$ flow rate during growth on the superconducting and dimensional properties of TiN films. We epitaxially grew TiN films on MgO substrates using reactive magnetron sputtering with varying $N_2$ flow rates and performed detailed measurements of their superconducting behavior. The results indicate that the superconducting critical temperature ($T_c$), residual resistivity ratio (RRR), normal-state resistivity ($\rho_n$), and upper critical magnetic field change significantly with $N_2$ flow rate. The films exhibit optimal quality at a flow rate of 1.5 sccm. Additionally, we observed anisotropy in the upper critical magnetic field. Importantly, we identified other significant factors influencing superconducting dimensionality beyond film thickness.

# 2 Experimental

We deposited TiN films on 10 mm×10 mm MgO substrates using DC reactive magnetron sputtering. The titanium target had a purity of 99.9999%, and the base vacuum of the reaction chamber was approximately $6\times10^{-8}$ Pa. The sputtering process occurred in a mixed atmosphere of high-purity Ar (99.9999%) and $N_2$ (99.9999%). During sputtering, the Ar flow rate was fixed at 15 sccm (1 sccm=1 mL/min@0 °C, $1.013\times10^5$ Pa), while the $N_2$ flow rate was adjusted within the range of 0.8-1.8 sccm. In this paper, samples are named according to the $N_2$ flow rate. For example, the sample grown at an $N_2$ flow rate of 1.3 sccm is designated as TiN_1.3. Other parameters remained constant, as detailed in Table 1. The film thickness was controlled to approximately 35 nm by adjusting the deposition time.

**Table 1** Growth parameters of TiN films.

| Parameter | Value |
|---|---|
| Ar flow rate/sccm | 15 |
| N2 flow rate/sccm | 0.8-1.8 |
| Power/W | 151 |
| Deposition pressure/Pa | 0.22 |
| Voltage/V | 430 |
| Current/A | 0.36 |
| Temperature/°C | 500 |

We characterized the crystal structure and surface morphology of the TiN thin films using an X-ray diffractometer (XRD, Bruker D8 diffractometer) and an atomic force microscope (AFM, Bruker Dimension Icon). We determined the film thickness by measuring X-ray reflectivity (XRR, Bruker D8 Discover). We measured electrical transport properties using a physical property measurement system (PPMS, Quantum Design) via the standard four-probe method. We applied the external magnetic field in two directions: perpendicular and parallel to the film surface.

# 3 Results and Analysis

## 3.1 Thin Film Characterization

We characterized the crystal structure of the obtained thin films using XRD. Figure 1(a) presents typical XRD patterns of TiN thin films grown at two different $N_2$ flow rates. Both films exhibited a distinct (200) diffraction peak at approximately 42.6°, confirming the successful growth of high-quality TiN thin films. We fitted the XRR data using a simple three-layer model (MgO substrate/TiN film/TiO layer) to obtain thickness information. Figure 1(b) displays the XRR data (black circles) and the corresponding fitting curve (orange line) for the representative sample TiN_1.3. The fitting results confirmed a thickness of approximately 35 nm for this sample. XRR tests on other samples indicated that all samples in this study possessed similar thicknesses.

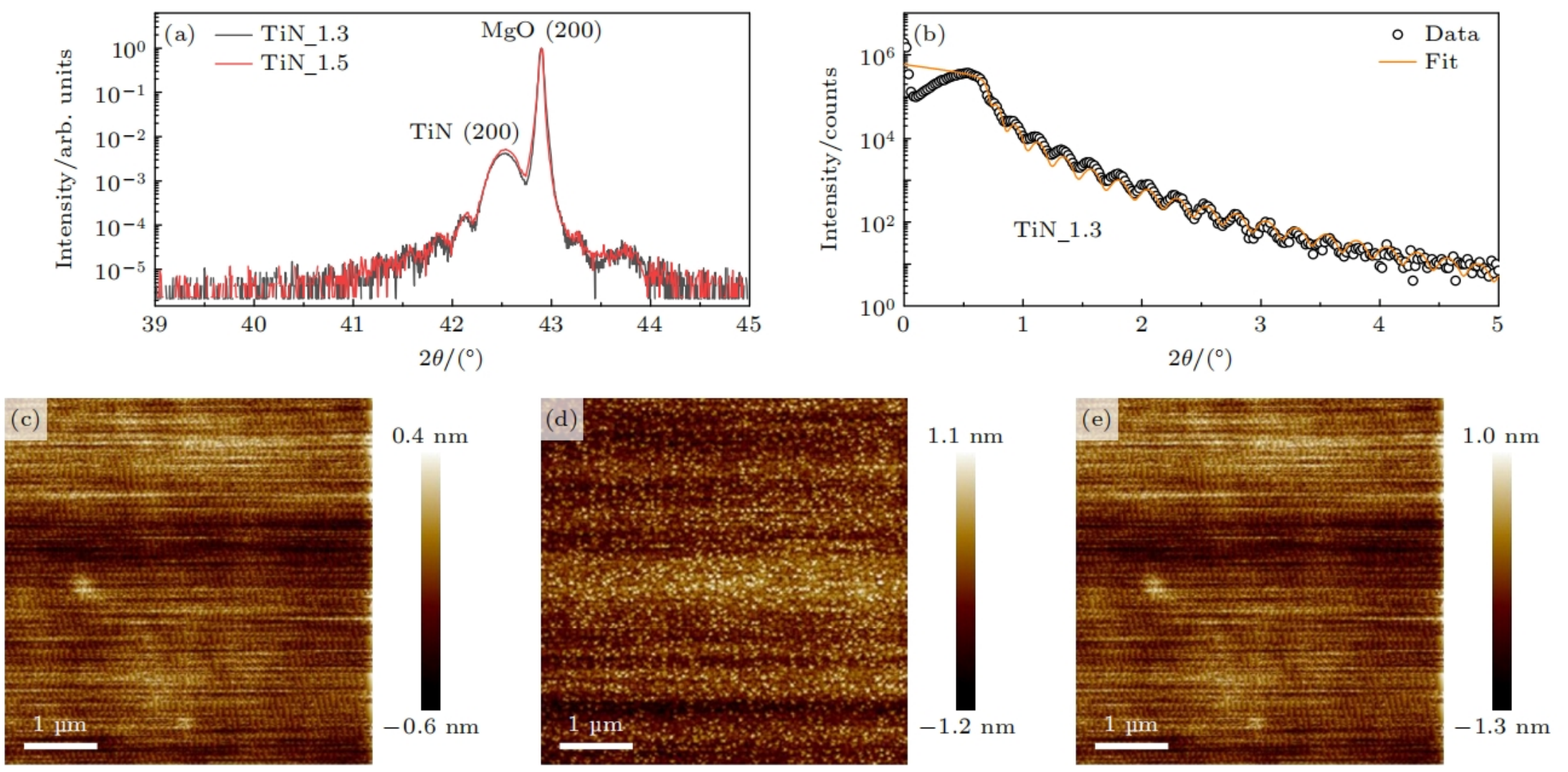

**Fig. 1 (a) X-ray diffraction patterns of TiN films grown under different $N_2$ flow rates; (b) the X-ray reflection curve of the TiN sample with $N_2$ flow rate of 1.3 sccm. The orange curve shows the fitting results; (c)–(e) AFM images of TiN thin films deposited at $N_2$ flow rates of 1.2, 1.5, and 1.6 sccm, respectively.**

Furthermore, we characterized the surface morphology of the TiN thin films using atomic force microscopy (AFM). Figures 1(c)—(e) present the AFM images of three representative samples: TiN_1.2, TiN_1.5, and TiN_1.6. As shown in Figures 1(c)—(e), all three films exhibit smooth surfaces within the 5 μm scan range. They feature uniformly distributed grains and low root-mean-square roughness ($R_q$). The measured $R_q$ values for these three samples are 0.142 nm, 0.325 nm, and 0.319 nm, respectively. The excellent surface flatness of the films can be attributed to the good lattice matching between TiN and the MgO substrate, as well as the successful control of growth conditions during the experiment.

## 3.2 Characteristics of the Superconducting and Normal States

We investigated the physical properties of samples grown at different $N_2$ flow rates in both the superconducting and normal states through electrical transport measurements. Figure 2(a) displays the temperature dependence of the normalized resistivity ($\rho/\rho_{300\,\mathrm{K}}$). The resistivity behavior in the normal state shows a systematic evolution with the $N_2$ flow rate. As the $N_2$ flow rate decreases, the metallic character of the TiN films in the normal state is gradually suppressed. When the $N_2$ flow rate drops to 0.8 sccm, the resistivity becomes nearly independent of temperature. In the low-temperature region, a sharp superconducting transition is observable in most samples ($N_2$ flow rate ≥ 1.1 sccm). Figure 2(b) presents a magnified view of the resistivity data in the low-temperature range, allowing for a clearer observation of the superconducting transition. As indicated in Figure 2(b), the $T_c$ of the TiN films gradually decreases as the $N_2$ flow rate is reduced. When the $N_2$ flow rate falls below 1.1 sccm, zero resistance disappears.

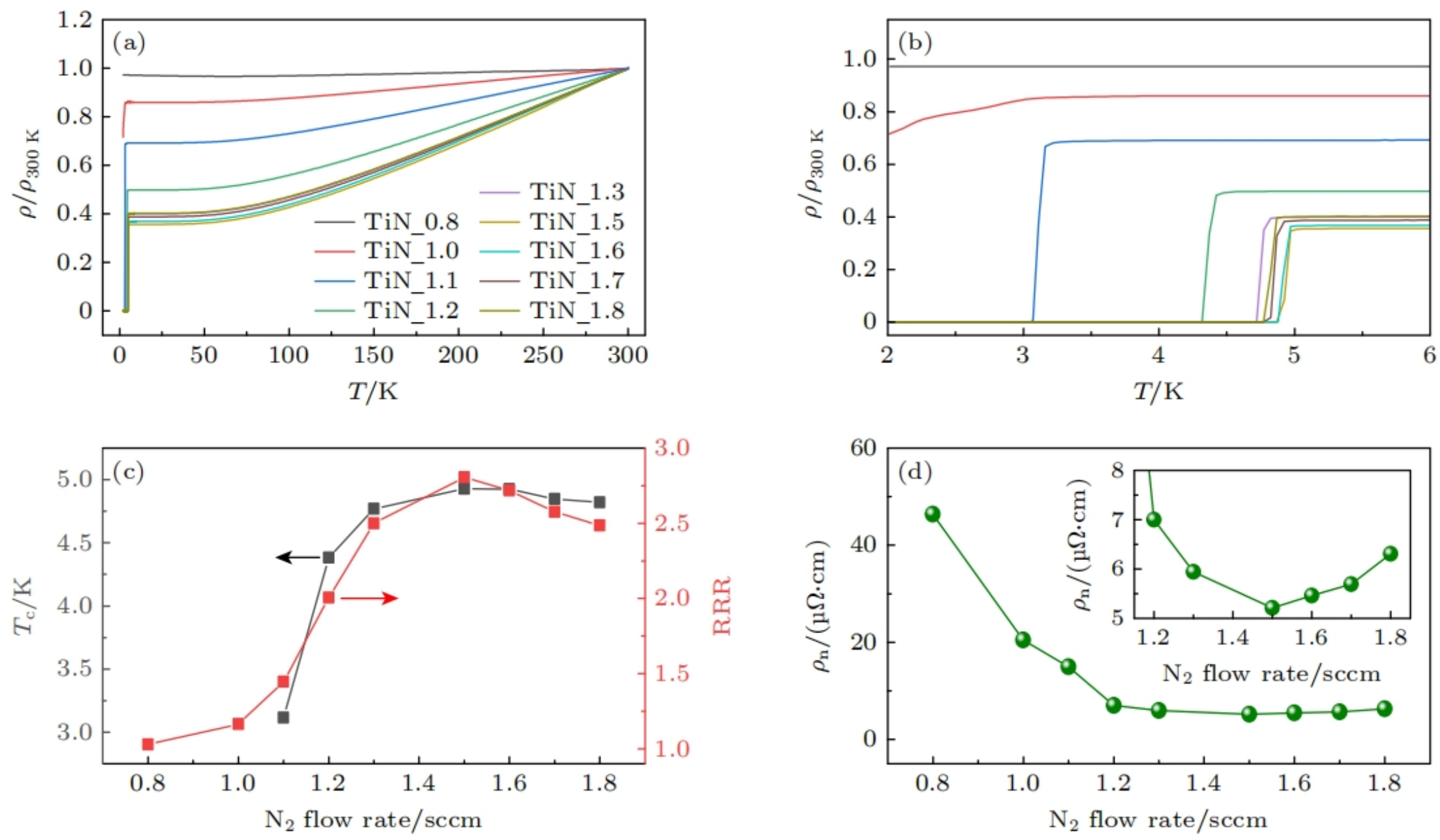

**Fig. 2 (a) Normalized resistivity of the samples grown under different $N_2$ flow rates under zero magnetic field varies with temperature; (b) an enlarged view of the normalized resistivity in the low-temperature region; (c) the SC critical temperature $T_c$ (left) and the residual resistivity ratio RRR (right) are functions of $N_2$ flow rate; (d) normal-state resistivity at 5 K ($\rho_n$) as a function of**

**$N_2$ flow rate, inset shows the enlarged view of the data in the range 1.2–1.8 sccm.**

Figure 2(c) summarizes the variations in $T_c$ and RRR with $N_2$ flow rate. Here, the $T_c$ value is determined using the 50% $\rho_n$ criterion, where $\rho_n$ represents the normal-state resistivity at 5 K above the superconducting transition. The RRR is defined as $\rho_{300\,K}/\rho_n$, which is the ratio of the resistivity at 300 K to the normal-state resistivity at low temperatures. As shown in Figure 2(c), $T_c$ increases significantly with the $N_2$ flow rate below approximately 1.5 sccm and subsequently saturates. In contrast, the RRR exhibits a trend of initial increase followed by a decrease. Additionally, we compared the normal-state resistivity $\rho_n$ of the samples at low temperatures, as illustrated in Figure 2(d). The results indicate that $\rho_n$ decreases markedly with increasing $N_2$ flow rate in the low-flow region, suggesting a reduction in electron scattering effects. To clarify the trend in the high-flow region, the data within the range of 1.2—1.8 sccm are presented in the inset of Figure 2(d). This inset reveals that $\rho_n$ reaches a minimum of approximately 5.2 μΩ·cm at an $N_2$ flow rate of 1.5 sccm. Clearly, the evolutionary trends of RRR and $\rho_n$ with respect to the $N_2$ flow rate corroborate each other. These consistent findings indicate that the quality of the TiN films is optimal at an $N_2$ flow rate of 1.5 sccm, under which condition the $T_c$ of the TiN film is 4.93 K.

It is noteworthy that the aforementioned results were obtained under high background vacuum (approximately $6\times10^{-8}$ Pa). During the experiments, we observed that the superconducting properties of the TiN films deteriorated significantly when the background vacuum level decreased. Figure 3 presents the resistivity data for two samples prepared at the same $N_2$ flow rate (1.1 sccm) but under different background vacuums. The samples TiN_1.1 and TiN_1.1-2 were grown under background pressures of $6\times10^{-8}$ Pa and $6\times10^{-6}$ Pa, respectively. The results show that when the background pressure increased to $6\times10^{-6}$ Pa, $T_c$ of the sample dropped from 3.10 K to 2.24 K. Simultaneously, the RRR value decreased from 1.45 to 1.23. This finding demonstrates that the background vacuum level in the sputtering chamber has a significant impact on the superconducting behavior of the prepared TiN films.

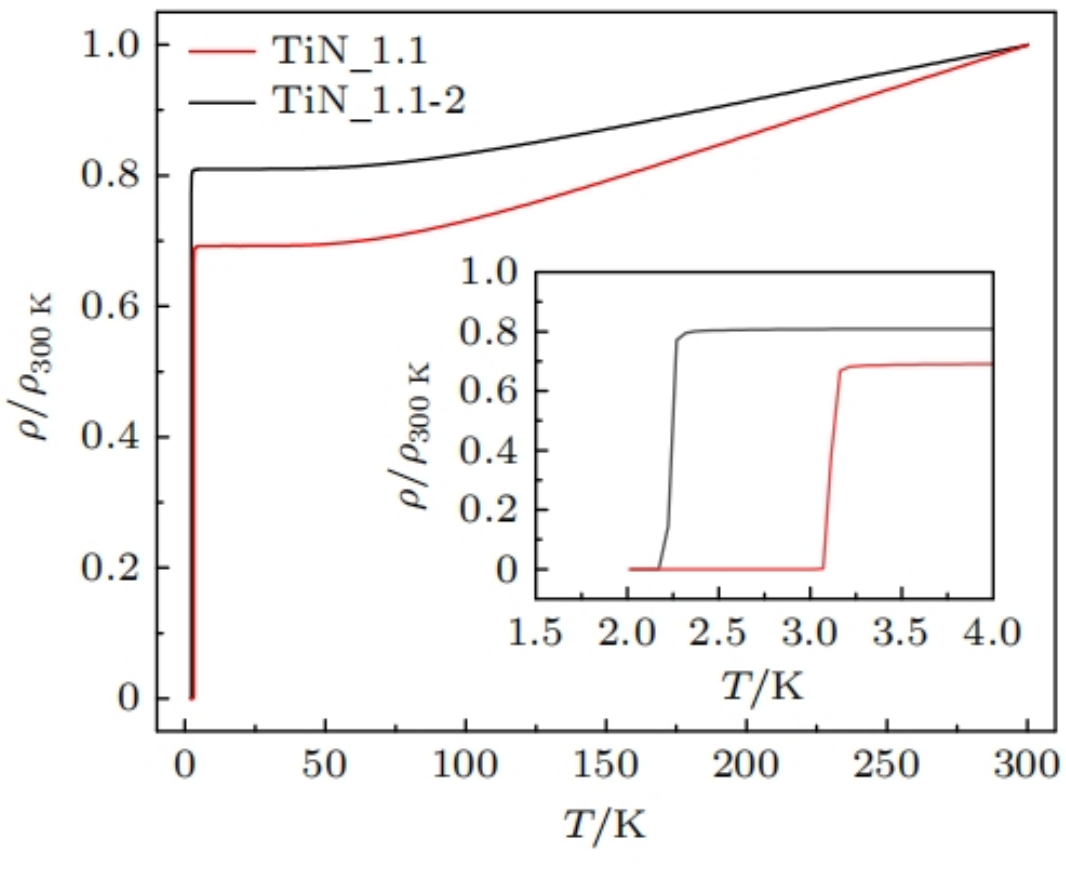

**Fig. 3 Temperature dependence of normalized resistivity of the samples grown with the $N_2$ flow rates of 1.1 sccm under different background vacuum levels. The inset shows an enlarged view of the data in the low-temperature region.**

### 3.3 Upper Critical Magnetic Field and Its Anisotropy

We systematically investigated the upper critical magnetic field behavior of TiN films by measuring resistivity-temperature data under magnetic fields of varying magnitudes and orientations. Figures 4(a)-(h) present the resistivity-temperature curves for four representative samples under magnetic fields applied perpendicular and parallel to the film surface. The data clearly indicate that the superconducting transition shifts to lower temperatures as the magnetic field increases. Using the 50% $\rho_n$ criterion, we determined the temperature dependence of the upper critical field $B_{c2}(T)$ for these four samples, as summarized in Figures 4(i)-(l). Notably, the upper critical field parallel to the film surface ($B_{c2,//}$) is significantly higher than that in the perpendicular direction ($B_{c2,\perp}$), indicating pronounced anisotropy in the TiN films. Furthermore, the temperature evolution of $B_{c2}(T)$ exhibits distinct characteristics for different field orientations: $B_{c2,\perp}(T)$ shows an approximately linear relationship with temperature when the field is perpendicular to the film surface. In contrast, $B_{c2,//}(T)$ displays negative curvature under parallel magnetic fields. For two-dimensional superconducting systems, the upper critical field in the parallel direction satisfies the relation [51] $B_{c2,//}(T) \propto (1-T/T_c)^{0.5}$. To evaluate the agreement of our experimental data with this relationship, we treated the exponent as a variable parameter $n$, yielding the following equation:

$$B_{c2,//}(T) = B_{c2,//}(0) \times (1 - T/T_c)^n, \qquad (1)$$

Here, $B_{c2,//}(0)$ and $n$ serve as fitting parameters. Clearly, when the parameter $n = 0.5$, this behavior aligns with the typical characteristics of two-dimensional superconductivity[51]. The term $B_{c2,//}(0)$ represents the value of $B_{c2,//}$ at 0 K. The fitting curves based on Equation (1) appear as black solid lines in Figures 4(i)—(l).

Figures 5(a) and 5(b) summarize the upper critical field data for all six samples. These figures reveal that the upper critical field exhibits clear evolution with $N_2$ flow rate for both magnetic field orientations. We first examine the anisotropy of $B_{c2}$, characterized by the anisotropy parameter $\Gamma = B_{c2,//}/B_{c2,\perp}$. As shown in Figure 5(c), the $\Gamma$ values for all six samples gradually decrease as temperature drops. This trend resembles observations in other superconducting thin films[8]. Notably, the $\Gamma$ value changes systematically with $N_2$ flow rate. Figure 5(d) compares the $\Gamma$ values of the samples at temperature $T = 0.8T_c$. The data indicate that $\Gamma$ increases linearly with $N_2$ flow rate, except for an anomalous region between 1.3 and 1.5 sccm where the upward trend suddenly slows.

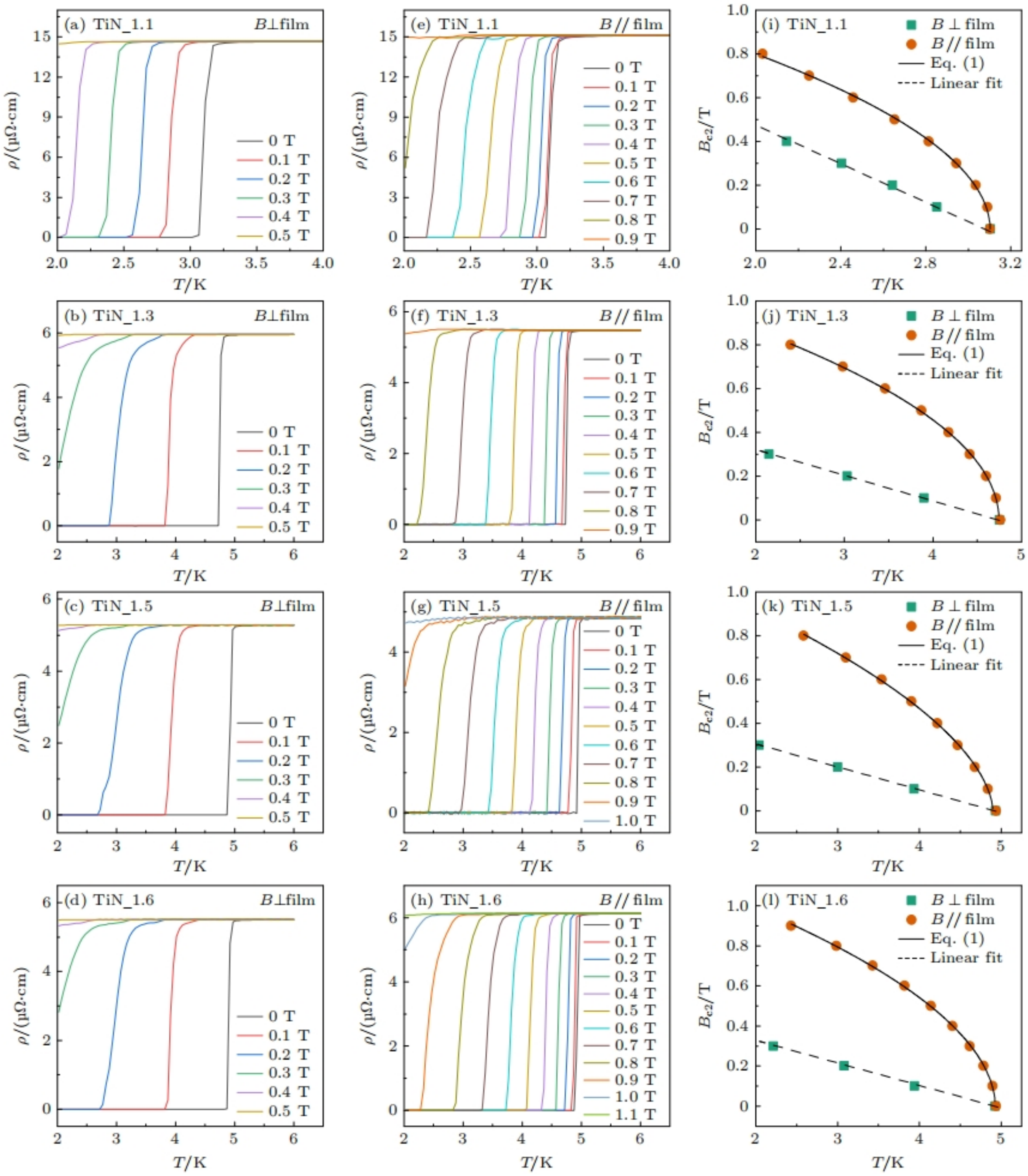

**Fig. 4 (a)–(d) Temperature dependence of resistivity of four samples under out-of-film field; (e)–(h) temperature dependence of resistivity of the same samples under in-film field; (i)–(l) the upper critical field $B_{c2}$ in two directions of the four samples varies with temperature.**

To facilitate quantitative comparison, we calculated the slope of the upper critical field in the vertical direction near $T_c$, defined as $B^* = -\,\mathrm{d}B_{c2,\perp}/\mathrm{d}T|_{T=T_c}$. The results are presented in Figure 5(e). The $B^*$ value exhibits a negative correlation with the $N_2$ flow rate over a broad range, reaching a minimum at 1.5 sccm. It is well established that this slope correlates closely with the normal-state resistivity[8], which reflects the extent of impurity scattering within the film. Consequently, the minimum $B^*$ observed at 1.5 sccm indicates that films grown under these conditions experience the lowest impurity scattering. This condition also yields the largest coherence length parallel to the film plane, $\xi(0)$, as shown in Figure 5(f). This finding aligns with conclusions drawn from $T_c$ and RRR data (see Figure 2(c)), further confirming that films grown at this flow rate possess optimal quality.

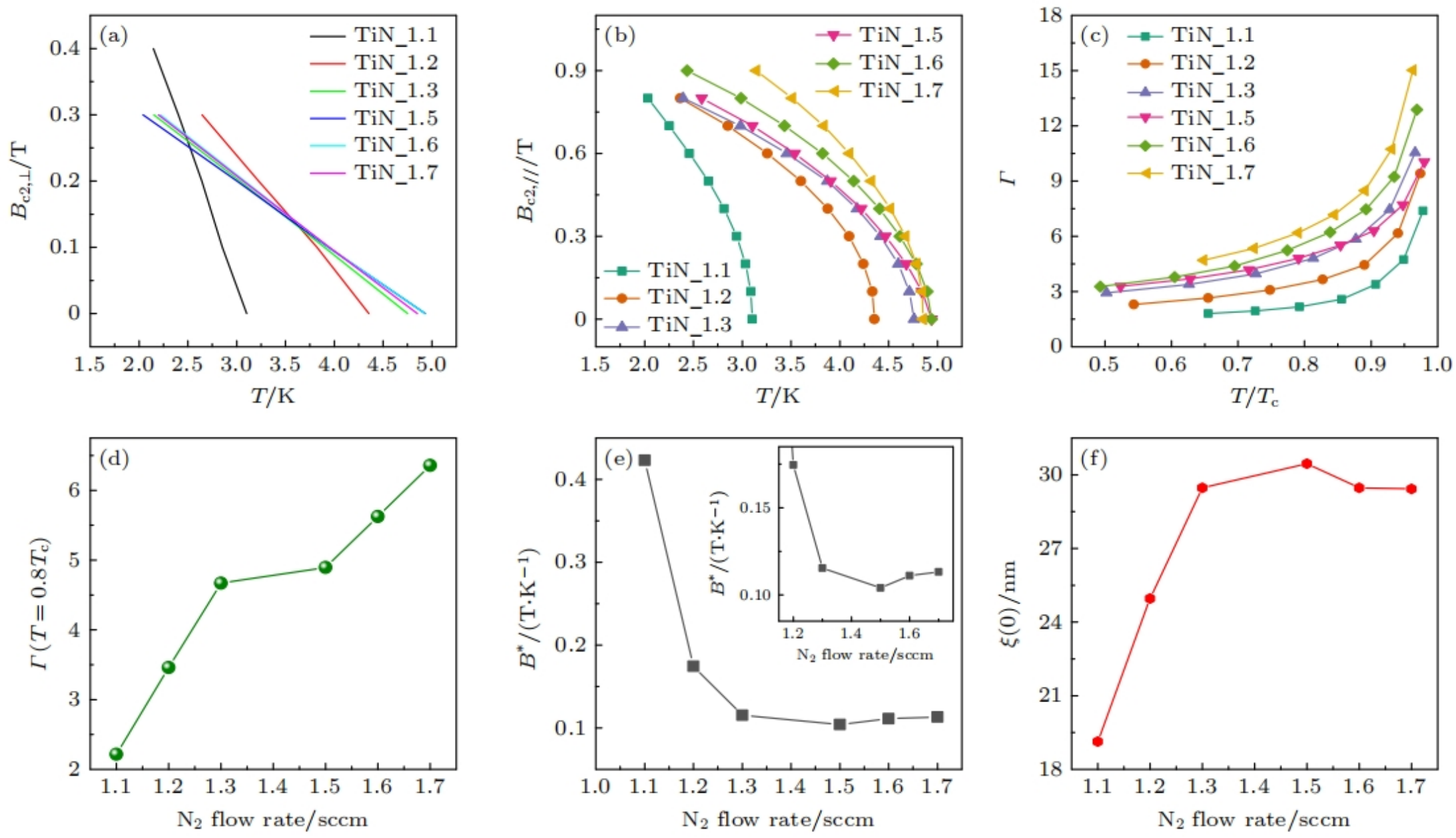

**Fig. 5 (a) , (b) Temperature dependence of the out-of-film and in-film upper critical field ($B_{c2,\perp}$ and $B_{c2,//}$) of the films with different $N_2$ flow rates; (c) anisotropic parameter $\Gamma$ as a function of the reduced temperature $T/T_c$; (d) values of Γ at the temperature $T = 0.8T_c$; (e) slope of $B_{c2,\perp}$ near $T_c$; (f) coherence length at zero temperature $\xi(0)$ as a functions of $N_2$ flow rate. The inset of (e) shows an enlarged view of the data.**

## 3.4 Dimensional Characteristics

The temperature evolution of the upper critical field in the parallel direction provides crucial insights into the dimensional characteristics of superconductivity[51]. To enable clearer comparison, we normalized the $B_{c2,//}(T)$ data for the six samples using $B_{c2,//}(0)$ and $T_c$. The results are displayed in Figure 6(a). The data for five samples, excluding TiN_1.5, generally follow a similar evolution pattern. Near $T/T_c = 1$, the ratio $B_{c2,//}(T)/B_{c2,//}(0)$ rises sharply. As the temperature decreases, this upward trend gradually flattens. In contrast, the data for sample TiN_1.5 exhibit a more gradual increase near $T/T_c = 1$, resulting in overall lower values compared to the other samples. The evolution characteristics of the $B_{c2}(T)/B_{c2,//}(0)$-$T/T_c$ curves can be quantitatively evaluated using the parameter $n$ in Eq. (1). Figure 6(b) illustrates the variation of $n$ values with $N_2$ flow rate. Clearly, the $n$ values for all samples are approximately 0.5, indicating two-dimensional superconducting characteristics in the films[51]. Notably, the $n$ value for sample TiN_1.5 is 0.58, which is significantly higher than that of other samples and exceeds the expected value of 0.5 for two-dimensional superconductivity. This suggests a deviation from two-dimensional superconductivity in this sample. To further verify the reliability of this behavior, we analyzed the data based on two additional criteria for calculating $B_{c2,//}$ (10%$\rho_n$ and 90%$\rho_n$). As shown in Figure 6(b), the evolution of $n$ values derived from these two criteria with $N_2$ flow rate exhibits a trend consistent with the data obtained using the 50%$\rho_n$ criterion. This indicates that the observed behavior is independent of the specific criterion used.

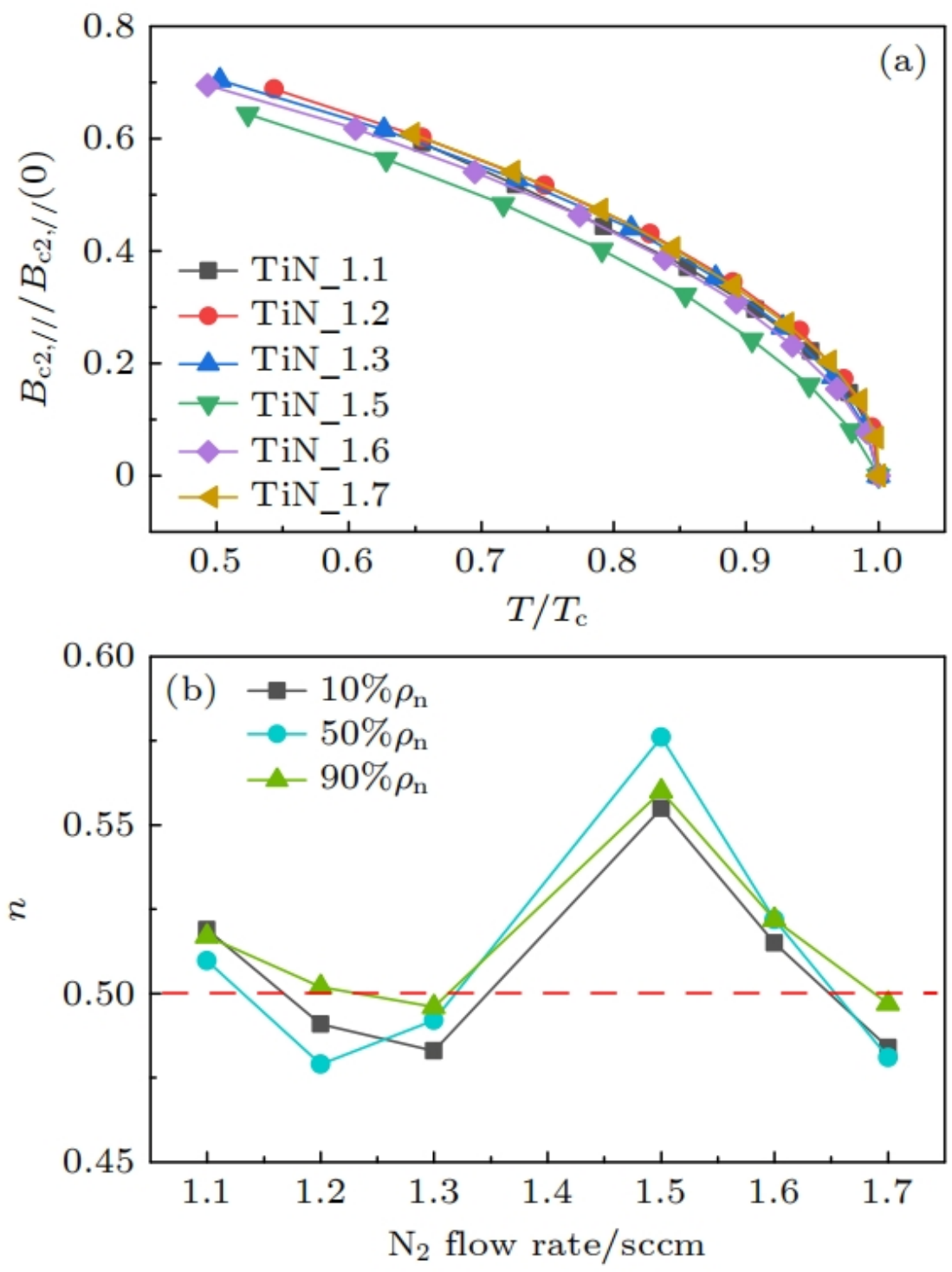

**Fig. 6 (a) Normalized $B_{c2,//}$ as a function of the reduced temperature $T/T_c$; (b) the fitting parameter $n$ of the $B_{c2,//}$ data based on three different criterion as a functions of $N_2$ flow rate.**

# 4 Discussion

From the perspective of device applications, superconducting performance and film flatness are the two most critical factors to consider. The data obtained in this study demonstrate that sample TiN_1.5 exhibits the best superconducting performance, as evidenced by its superior superconducting critical temperature, residual resistivity ratio (RRR), and upper critical magnetic field behavior.

Regarding surface morphology, our results reveal characteristics distinctly different from those of TiN films grown on Si substrates. Previous experimental results on Si substrates[52] indicated that $N_2$ flow rate significantly affects the surface roughness of TiN films. Specifically, when the $N_2$ flow rate is less than 1.6 sccm, the films exhibit an island growth mode, leading to a significant increase in surface roughness. This contrasts sharply with the results obtained on MgO substrates in this study. As shown in Figures 1(c)-(e), samples obtained at $N_2$ flow rates as low as 1.2 sccm still display very flat surfaces. This result suggests that the good lattice matching between MgO and TiN suppresses the emergence of the island growth mode at low $N_2$ flow rates.

Given the significant influence of the substrate on TiN film growth, it is important to compare the superconducting properties of samples grown on different substrates. Table 2 summarizes the key superconducting parameters of optimal TiN films with the same thickness (35 nm) grown on three different substrates[28]. As indicated in Table 2, TiN samples grown on MgO substrates exhibit the highest RRR and the lowest $B^*$ values, further confirming the high quality of TiN films on MgO. However, their $T_c$ is slightly lower than that of samples on Si substrates. This difference may arise from the enhancement of superconductivity in the latter due to stress induced by lattice mismatch between the Si substrate and TiN.

**Table 2** Comparison of physical parameters of TiN films grown on different substrates.

| Substrate | $T_c$/K | RRR | $B^*/(T\cdot K^{-1})$ | Reference |
|---|---|---|---|---|
| Sapphire | 4.71 | 1.55 | 0.46 | [28] |
| Si | 5.14 | 2.09 | 0.33 | [28] |
| MgO | 4.93 | 2.81 | 0.10 | This work |

The deviation from two-dimensional superconductivity observed in the highest-quality TiN_1.5 sample is a noteworthy issue. Typically, studies on dimensional transitions in thin-film superconductivity focus primarily on the relationship between film thickness and the superconducting coherence length. When the film thickness decreases to a value smaller than or comparable to the coherence length, the superconducting behavior transitions from three-dimensional to two-dimensional. In this study, the thickness of all samples was fixed at 35 nm. Furthermore, our calculations indicate that the vertical coherence length of the TiN_1.5 sample is very close to that of the TiN_1.6 sample. However, the $n$ value of TiN_1.5 is significantly higher than that of the other samples. Therefore, our results clearly demonstrate that factors other than film thickness are at play. We propose that this finding provides a paradigm for investigating how internal states of thin films influence their superfluid behavior. It is well established that thermal fluctuations and superfluid stiffness are two key factors determining two-dimensional superconducting behavior[53]. Systems exhibiting two-dimensional superconductivity are typically accompanied by significant thermal fluctuation effects and a corresponding reduction in superfluid stiffness. Conversely, the presence of disorder and defects reduces the superfluid stiffness of the material. In this sense, high quality implies high superfluid stiffness, which helps resist the influence of fluctuations and makes the system behavior more akin to three-dimensional characteristics, albeit to a lesser extent.

## 5 Conclusion

In this study, high-quality TiN superconducting thin films were successfully grown on MgO(100) substrates using DC reactive magnetron sputtering. Experimental results indicate that the $N_2$ flow rate significantly affects the crystalline quality and superconducting properties of the films. The film grown at an $N_2$ flow rate of 1.5 sccm exhibited the best overall performance, including a high $T_c$, a large RRR, a low $\rho_n$, and an extremely low slope of the upper critical field with respect to temperature in the vertical direction. Unlike the case with Si substrates, the surface roughness of TiN films on MgO substrates was largely unaffected by the $N_2$ flow rate. The upper critical magnetic field displayed significant anisotropic behavior. It is worth emphasizing that our results reveal other important factors influencing the dimensional characteristics of superconducting thin films beyond film thickness. This study provides an important experimental basis for optimizing the quality of TiN films and offers new perspectives for investigating dimensional effects in low-dimensional superconducting systems.

## References

[1] Haviland D B, Liu Y, Goldman A M 1989 *Phys. Rev. Lett.* **62** 2180

[2] Saito Y, Nojima T, Iwasa Y 2017 *Nat. Rev. Mater.* **2** 16094

[3] Chen J X, Li H 2025 *J. Funct. Mater. Devices* **31** 1

[4] Zhou G D, Lü W, Wang H, Nie Z H, Chen Y Q, Li Y Y, Huang H L, Chen W Q, Sun Y J, Xue Q K, Chen Z Y 2025 *Nature* **640** 641

[5] Fang Y Q, Tu X Y, Liu Y X, Peng W, Mu G, Huang F Q 2025 *J. Funct. Mater. Devices* **31** 42

[6] Xu L, Zhang X F, You L X 2025 *Acta Phys. Sin.* **74** 187402

[7] Wang Z, Kawakami A, Uzawa Y, Komiyama B 1996 *J. Appl. Phys.* **79** 7837

[8] Wang X N, Wang L J, Liu Y X, Gao W P, Wu Y, Xu Z L, Jin H, Zhang L, Peng W, Wang Z, Li W, Mu G, Lin Z R 2023 *Physica C* **606** 1354223

[9] Chen Y, Zheng X Q, Zhou H, Chen J X, Ma M, Li H, Zhang X F, You L X 2025 *J. Funct. Mater. Devices* **31** 486

[10] Vissers M R, Gao J, Wisbey D S, Hite D A, Tsuei C C, Corcoles A D, Steffen M, Pappas D P 2010 *Appl. Phys. Lett.* **97** 232509

[11] Diener P, Schellevis H, Baselmans J J A 2012 *Appl. Phys. Lett.* **101** 252601

[12] Oliver W D, Welander P B 2013 *MRS Bull.* **38** 816

[13] Kim S, Terai H, Yamashita T, Qiu W, Fuse T, Yoshihara F, Ashhab S, Inomata K, Semba K 2021 *Commun. Mater.* **2** 98

[14] Gao J , Vissers M R, Sandberg M O, Da Silva F C S, Nam S W, Pappas D P, Wisbey D S, Langman E C, Meeker S R, Mazin B A, Leduc H G, Zmuidzinas J,

Irwin K D 2012 *Appl. Phys. Lett.* **101** 142602

[15] Ho Eom B, Day P K, LeDuc H G, Zmuidzinas J 2012 *Nat. Phys.* **8** 623

[16] Longden J, Chaumont C, Klimovich N, Wood S, Boussaha F, Tan B 2025 *Open Research Europe* **5** 109

[17] Yadav S, Kaushik V, Saravanan M P, Aloysius R P, Ganesan V, Sahoo S 2021 *Sci. Rep.* **11** 7888

[18] Chang J B, Vissers M R, Córcoles A D, Sandberg M, Gao J, Abraham D W, Chow J M, Gambetta J M, Beth Rothwell M, Keefe G A, Steffen M, Pappas D P 2013 *Appl. Phys. Lett.* **103** 012602

[19] Jaim H I, Aguilar J, Sarabi B, Rosen Y, Ramanayaka A, Lock E, Richardson C, Osborn K 2014 *IEEE Trans. Appl. Supercond.* **25** 1

[20] Torgovkin A, Chaudhuri S, Ruhtinas A, Lahtinen M, Sajavaara T, Maasilta I J 2018 *Supercond. Sci. Technol.* **31** 055017

[21] Shearrow A, Koolstra G, Whiteley S J, Earnest N, Barry P S, Heremans F J, Awschalom D D, Shirokoff E, Schuster D I 2018 *Appl. Phys. Lett.* **113** 212601

[22] Narayan J, Tiwari P, Chen X, Singh J, Chowdhury R, Zheleva T 1992 *Appl. phys. Lett.* **61** 1290

[23] Vispute R D, Talyansky V, Sharma R P, Choopun S, Downes M, Venkatesan T, Li Y, Salamanca-Riba L, Iliadis A A, Jones K A, Iliadis A A, Asif Khan M, Yang J W 1998 *Appl. Sur. Sci.* **127** 431

[24] Krockenberger Y, Karimoto S, Yamamoto H, Semba K 2012 *J. Appl. Phys.* **112** 083920

[25] Hadacek N, Sanquer M, Villégier J -C 2004 *Phys. Rev. B* **69** 024505

[26] Baturina T I, Islamov D R, Bentner J, Strunk C, Baklanov M R, Satta A 2004 *J. Exp. Theor. Phys. Lett.* **79** 337

[27] Mironov A Y, Baturina T I, Vinokur V M, Postolova S V, Kropotin P N, Baklanov M R, Nasimov D A, Latyshev A V 2010 *Physica C* **470** S808

[28] Liu Y X, Liu Y C, Xu Z L, Yu A B, Wang X N, Peng W, Wu Y, Mu G, Lin Z R 2024 *Supercond. Sci. Technol.* **37** 105015

[29] Ohya S, Chiaro B, Megrant A, Neill C, Barends R, Chen Y, Kelly J, Low D, Mutus J, OMalley P J J, Roushan P, Sank D, Vainsencher A, Wenner J, White T C, Yin Y, Schultz B D, Palmstrøm C J, Mazin B A, Cleland A N, Martinis J M 2013 *Supercond. Sci. Technol.* **27** 015009

[30] Bollinger A T, Dubuis G, Yoon J, Pavuna D, Misewich J, Božović I 2011 *Nature* **472** 458

[31] Leng X, Garcia-Barriocanal J, Bose S, Lee Y, Goldman A M 2011 *Phys. Rev. Lett.* **107** 027001

[32] Wang F, Biscaras J, Erb A, Shukla A 2021 *Nat. Commun.* **12** 2926
[33] Wang T, Yu A B, Liu Y X, Gu G D, Peng W, Di Z F, Jiang D, Mu G 2022 *Phys. Rev. B* **106** 104509
[34] Zaikin A D, Golubev D S, van Otterlo A, Zimányi G T 1997 *Phys. Rev. Lett.* **78** 1552
[35] Giordano N 1998 *Phys. Rev. Lett.* **61** 2137
[36] Sharifi F, Herzog A V, Dynes R C 1993 *Phys. Rev. Lett.* **71** 428
[37] Lau C N, Markovic N, Bockrath M, Bezryadin A, Tinkham M 2001 *Phys. Rev. Lett.* **87** 217003
[38] Wang X N, Xu Z L, Liu X Y, Xu F, Liu Y X, Gao W P, Wu Y, Yang M, Peng W, Wang Z 2024 *Supercond. Sci. Technol.* **37** 045001
[39] Xing Y, Zhang H M, Fu H L, Liu H W, Sun Y, Peng J P, Wang F, Lin X, Ma X C, Xue Q K, Jian W, Xie X C 2015 *Science* **350** 542
[40] Saito Y, Nojima T, Iwasa Y 2018 *Nat. Commun.* **9** 778
[41] Han X W, Wu Y, Xiao H, Zhang M, Gao M, Liu Y, Wang J, Hu T, Xie X M, Di Z F 2020 *Adv. Sci.* **7** 1902849
[42] Liu Y, Qi S C, Fang J C, Sun J, Liu C, Liu Y Z, Qi J J, Xing Y, Liu H W, Lin X, Wang L L, Xue Q K, Xie X C, Wang J 2021 *Phys. Rev. Lett.* **127** 137001
[43] Wang X N, Wang L J, Liu Y X, Chen F, Gao W P, Wu Y, Xu Z L, Peng W, Wang Z, Di Z F, Li W, Mu G, Lin Z R 2023 *Phys. Rev. B* **107** 094509
[44] De la Barrera S C, Sinko M R, Gopalan D P, Sivadas N, Seyler K L, Watanabe K, T. Taniguchi, Tsen A W, Xu X D, Xiao D, Hunt B M 2018 *Nat. Commun.* **9** 1427
[45] Xi X X, Wang Z F, Zhao W W, Park J -H , Law K T, H. Berger, Forro L, Shan J, Mak K F 2016 *Nat. Phys*. **12** 139
[46] Lu J M, Zheliuk O, Leermakers I, Yuan N F Q, Zeitler U, Law K T, Ye J T 2015 *Science* **350** 1353
[47] Tsen A W, Hunt B, Kim Y D, Yuan Z J, Jia S, Cava R J, Hone J, Kim P, Dean C R, Pasupathy A N 2016 *Nat. Phys.* **12** 208
[48] Phillips P, Dalidovich D 2003 *Science* **302** 243
[49] Ephron D, Yazdani A, Kapitulnik A, Beasley M R 1996 *Phys. Rev. Lett.* **76** 1529
[50] Saito Y, Kasahara Y, Ye J, Iwasa Y, Nojima T 2015 *Science* **350** 409
[51] Tinkham M 1996 *Introduction to Superconductivity* (New York: McGraw-Hill
[52] Wu Y, Liu Y X, Gao W P, Wang X N, He X L, Jin H, Peng W, Mu G, Lin Z R, Wang Z 2024 *Supercond. Sci. Technol.* **37** 115002
[53] Beasley M R, Mooij J E, Orlando T P 1979 *Phys. Rev. Lett.* **42** 1165